\documentclass[journal]{IEEEtran}
\usepackage{soul, color}
\sethlcolor{yellow} 
\usepackage{graphicx}
\usepackage[justification=centering]{caption}
\UseRawInputEncoding
\usepackage{bm}
\usepackage{float}

\usepackage{ifpdf}

\usepackage{cite}

\ifCLASSINFOpdf
  
\else
\fi
\usepackage{amsmath}
\usepackage{subfig}
\usepackage{cite}
\usepackage{times}

\usepackage{graphicx}

\makeatletter
\newcommand*\bigcdot{\mathpalette\bigcdot@{.5}}
\newcommand*\bigcdot@[2]{\mathbin{\vcenter{\hbox{\scalebox{#2}{$\m@th#1\bullet$}}}}}
\makeatother

\begin{document}
%
\title{Disaggregated Prefill and Decoding Inference System for Large Language Model Serving on Multi-Vendor GPUs}
%
%
%

\author{
    \IEEEauthorblockN{
        Xing Chen, Rong Shi, Lu Zhao, LingBin~Wang, Xiao~Jin, Yueqiang~Chen, Hongfeng~Sun*\\
    }
    \IEEEauthorblockA{
        \textit{ZTE Corporation, China}\\
    }

\thanks{Xing Chen, Rong Shi, Lu Zhao, Lingbin Wang, Xiao Jin, Yueqiang Chen, Hongfeng Sun are with Wireless and Computing Product R\&D Institute, ZTE Corporation, China (e-mail: chen.xing14@zte.com.cn; shi.rong@zte.com.cn; zhao.lu@zte.com.cn; wang.lingbin@zte.com.cn; jin.xiao5@zte.com.cn; chen.yueqiang@zte.com.cn; sun.hongfeng@zte.com.cn; Corresponding author: Xing Chen, Hongfeng Sun)}
}

\maketitle

\begin{abstract}
LLM-based applications have been widely used in various industries, but with the increasing of models size, an efficient large language model (LLM) inference system is an urgent problem to be solved for service providers. Since the inference system is divided into two stage with different characteristics: Prefill and Decode, the two stage will interfere with each other during the inference process. Toward this end, a P-D disaggregated inference framework is proposed by some researchers. Current research is done on homogeneous GPUs, and lacks deployment solutions based on business scenarios. Compared with homogeneous GPUs, using heterogeneous GPUs to construct inference systems can better improve resource utilization and reduce costs. Even if GPUs from different vendors are used to build inference systems, on the basis of reducing costs, the resource utilization rate can be improved and the dependence on a single vendor can be reduced. Therefore, a P-D disaggreagetd inference system based on heterogeneous GPUs is designed, and the heterogeneous compatible transmission module in the system is designed to address heterogeneous GPU data compatibility issues. Then, a joint optimization algorithm of parallel strategy and instance number allocation is proposed to obtain the deployment solutions. Finally, the experimental results show that the P-D disaggregated inference system can well solve the hybrid inference problem of heterogeneous GPUs from different vendors, and the joint optimization algorithm can obtain the optimal deployment solution.
\end{abstract}


%
\IEEEpeerreviewmaketitle

\section{Introduction}
%
%
%
%
\IEEEPARstart{W}{ith} the full development of the Large Language Model (LLM) research, it has been applied to various industries with its powerful natural language interaction capabilities\cite{xu2024enhancing}. The user experience of LLM applications consists of two indicators: inference accuracy and inference speed\cite{wang2024understanding,feng2025canvil}. Inference accuracy depends on the training process of the LLM, while inference speed depends on the design of LLM inference system. In this paper, we mainly focus on the design of inference system. Recently, the number of LLM parameters has continued to increase, which has put forward higher performance requirements for LLM inference systems.

Based on the above analysis, how to improve the performance of inference system becomes a key problem for LLM applications\cite{stojkovic2024dynamollm,li2024llm}. The inference process of LLM inference system is divided into two stages. The first stage is the prefill stage, which generates the first token and is computationally intensive. The second stage is the decode stage, which generates the remaining tokens and is VRAM intensive\cite{jin2024p,chen2024kvdirect,hu2024inference}. In the previous inference system, the inference process adopted a prefill priority strategy\cite{kwon2023efficient}. During the inference process, as long as a new request arrives, the decoding operation being executed needs to be stopped and the prefill operation is executed first. The strategy will increase the decoding delay, and the differentiated resource requirement of the two stages will constrain each other during the inference process, resulting in reduced system throughput. In order to reduce this constraint, Chunked prefill was proposed\cite{agrawal2023sarathi}.

Chunked prefill allows building multiple decode max batches from a single Prefill request, maximizing coverage of piggybackable decodes. Therefore, the P-D disaggregated inference system architecture is proposed\cite{qin2024mooncake,qin2025mooncake}. The main feature of P-D disaggregated architecture is to deploy two identical model instances. One model instance is used as the prefill instance to complete the inference of the first token, another model instance is used as decode instance to complete the inference of the remaining tokens. The Remote Direct Memory Access(RDMA) technology is used to transmit the KV data between prefill instances and decode instances. In this way, the interference between the two inference stage can be reduced. Meanwhile, since the two stages are independently deployed, GPUs with different characteristics can be used to deploy prefill instance and decode instance, respectively. GPUs with strong computing capabilities are used to deploy prefill instances, and GPUs with strong memory access capabilities are used to deploy decode instances. Based on the above analysis, the throughput of the two instances is different. To improve resource utilization, it is necessary to ensure that the throughput of the prefill instance matches the throughput of the decode instance. Therefore, deciding the number of P instances and D instances is a key problem in the P-D disaggregated inference system.

Although the P-D disaggregated architecture reduces the interference between the prefill stage and decode stage, if the prefill stage and decode stage adopt the same GPU, the system resource utilization will be decreased. Thus, the P-D disaggregated inference system based on heterogeneous GPUs is proposed to save costs and improve resource utilization. Currently, heterogeneous GPU inference system is mainly aimed at different GPUs from the same vendor. However, with the continuous development of domestic GPUs, heterogeneous GPUs inference system from different vendors will receive extensive attention, which will increase the utilization rate of idle GPUs and reduce dependence on a single vendor.Therefore, how to design heterogeneous GPUs inference system from different manufactures is a core problem in the evolution of inference systems.

\IEEEpubidadjcol
In this paper, we focus on the design of P-D disaggregated inference system and the joint optimization problem of parallel strategy and instance ratio allocation for heterogeneous GPUs from different vendors. The heterogeneous P-D disaggregated inference system can maximize the utilization of GPU and save the system costs. The joint optimization problem can be divided into two steps. The first step is to jointly optimize the parallel strategy and number of P instances. The second step is to use the output of the first step as input to jointly optimize the parallel strategy and number of D instances. The contributions of this paper can be summarized as follows:
\begin{enumerate}
\item Heterogeneous compatible module design. If the GPU data of one vendor is to be used on the GPU of another vendor, a heterogeneous compatible module is designed. The module is mainly consisted of several components: precision alignment component, VRAM management alignment component (block size and layout) and parallel strategy alignment component.
\item Parallel strategy and instance ratio allocation algorithm. A parallel strategy and instance number allocation joint optimization problem of P-D disaggregated inference system is formulated for heterogeneous GPUs from different vendors. The problem is a serial two-stage optimization problem. The first stage is a optimization problem of parallel strategy, with the objective of maximize the throughput of each GPU, subject to Time to First Token (TTFT) and High Bandwidth VRAM (VRAM) of each GPU. The second stage is a joint optimization problem of parallel strategy and number of Decode instances, with the objective of maximize the throughput of each instance, subject to (TPOT) and VRAM of each GPU. These two problem are solved by the global search algorithm.
\item Detailed simulator design. A simulator is design to model the performance of the inference system based on the hardware capabilities, framework feature and different parallel strategy. During the modeling process, we modeled the inference system layer by layer, starting from the original transformer structure, and then considering the hardware characteristics and framework features to form the computing operator library and the communication operator library. Finally, the VRAM usage and latency were modeled to assist in optimizing algorithm.
\end{enumerate}

The rest of this paper is organized as follows: Section II reviews the development of the related work. Section III designs the P-D disaggregated inference system for heterogeneous GPUs from different vendors. Section IV models the joint optimization problem of parallel strategy and instance ratio allocation. The joint optimization algorithm based on the global search is proposed. Section V provides the simulation result and evaluates the performance of the proposed algorithm. Finally, Section VI concludes the paper.

\section{Related work}
Large language model have attracted widespread attention due to their outstanding performance in various tasks. However, the substantial computing and VRAM requirements of LLM inference system in the resource constrained scenarios pose a great challenge\cite{li2024llm,zhang2024edgeshard}. The industry committed to researching efficient inference system to improve user experience quality. Currently, several optimizations at the framework level are introduced as follows.

\subsection{Parallel Strategy}
Recently, the number of model parameters has increased, resulting in a single GPU or a single server being unable to accommodate these larger models. Therefore, GPipe\cite{huang2019gpipe}, a pipeline parallelism library is proposed by pipelining different sub-sequences of layers on separate accelerators. Next, Megatron-LM\cite{shoeybi2019megatron}, a efficient intra-layer model parallel approach is proposed to train transformer models with billions of parameters. The approach is orthogonal and complimentary to pipeline model parallelism, and can be inserted into the original PyTorch by implementing some communication operations. Simply using basic parallel methods can be to resource waste in large-scale model training. In \cite{narayanan2021efficient}, tensor, pipeline and data parallelism can be efficiently composed to train large-scale model using thousands of GPUs, and a novel interleaved pipelining schedule was proposed, resulting in a 10$\%$ increase in throughput. Since the model training takes up more GPU VRAM, these parallel methods were initially used in the training process of LLM. However, as the number of model parameters increases, a single GPU or a single node cannot accommodate the model parameters in the inference system. Therefore, these parallel methods are applied to the inference system. 

\subsection{P-D Disaggregated Inference System}
In general, the inference process is divided into two stages: Prefill and Decode. The Prefill stage is computationally intensive, and the decode stage is VRAM intensive. Due to their differentiated characteristics, these two stages have mutual constraints in the inference process. Initially, due to the low efficiency of inter-node communication, the disaggregated inference architecture was restricted to s single node, which limited the flexible resource scheduling and service capabilities. Therefore, KVDirect was developed by ByteDance\cite{chen2024kvdirect}, which can optimize the KV cache transfer in the disaggreagted LLM inference system. Mooncake\cite{qin2024kimi,qin2024mooncake}, a KVCache-centric disaggregated architecture that separates the prefill and decode clusters was proposed. The architecture also takes advantage of underutilized resources such as CPU, DRAM and SSD resources of GPU clusters to implement a disaggregated KV Cache. Experiments show that Mooncake achieves a 525$\%$ throughput gain under SLO constraints. At the same time, som works has more detailed algorithm research, such as kv transmission optimization\cite{li2025flowkv}, efficient scheduling\cite{wu2025arrow}. The above P-D disaggregated research adopts the same GPU to deply the inference system. However, this deployment method will waste a lot of GPU resources. Specifically, the comute-intensive prefill instance has low VRAM utilization, and the VRAM-intensive decode instance has low computing utilization. Liang\cite{liang2025injecting} also verifies this situation. 

\subsection{Hybrid Inference on Different GPUs from the Same vendors}
Based on the above analysis, the P-D disaggregated system of heterogeneous GPUs can better improve GPU utilization and save costs. Prefill instances are deployed with GPUs that have strong computing capabilities but weak VRAM access capabilities, and decode instances are deployed with GPUs that have strong VRAM acess acpabilities but weak computing capabilities. HexGen-2\cite{zhao2024hetegen,jiang2025hexgen}, a P-D disaggregated system on heterogeneous GPUs was constructed and a scheduling algorithm for computing and communication resource allocation was proposed to minimize costs. The experimental setup of HexGen-2 utilize four types of GPUs: H100, A100, L40, and A6000, to construct a heterogeneous GPUs cluster. However, the GPUs in the experimental are produced by the same vendor. In order to reduce the idle rate of resources and the dependence on a single vendor, the P-D disaggreagetd system of heterogeneous GPUs from different vendors is an urgent problem to be solved.

\section{System Design}
\begin{figure}
	\centering
	\includegraphics[width=8cm]{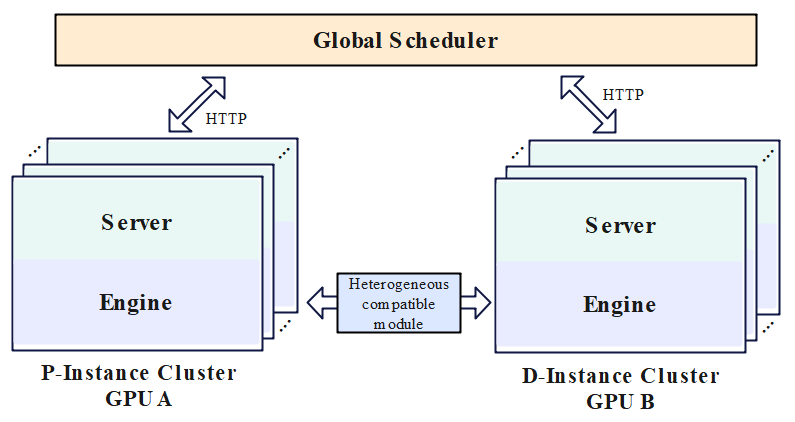}
	\caption{System architecture}
	\label{Fig.1}
\end{figure}

In this paper, we consider a hybrid inference scenario that leverages GPUs from different vendors in order to maximize resource utilization. As shown in Fig.\ref{Fig.1}, a P-D disaggregated inference system for heterogeneous GPUs from different vendors is established. The system consists of the following modules: (a) global scheduler, which can receive HTTP requests from users and forwards them to server module; (b) server, which can receive request from the global scheduler and add requests to the corresponding engine queue; (c) engine, which can execute the inference tasks and return the generated results to server module; (d) heterogeneous compatible module, which is mainly consisted of two components: VRAM management alignment component (block size and layout) and parallel strategy alignment component; (c) KV transfer module, which can provide the RDMA transmission function for KV data between P instance and D instance. In the hybrid inference scenario, GPUs with stronger computing capabilities are used as P instances, and GPUs with stronger VRAM access capabilities are used as D instances. 

\subsection{Workflow Design}
\begin{figure}
	\centering
	\includegraphics[width=8cm]{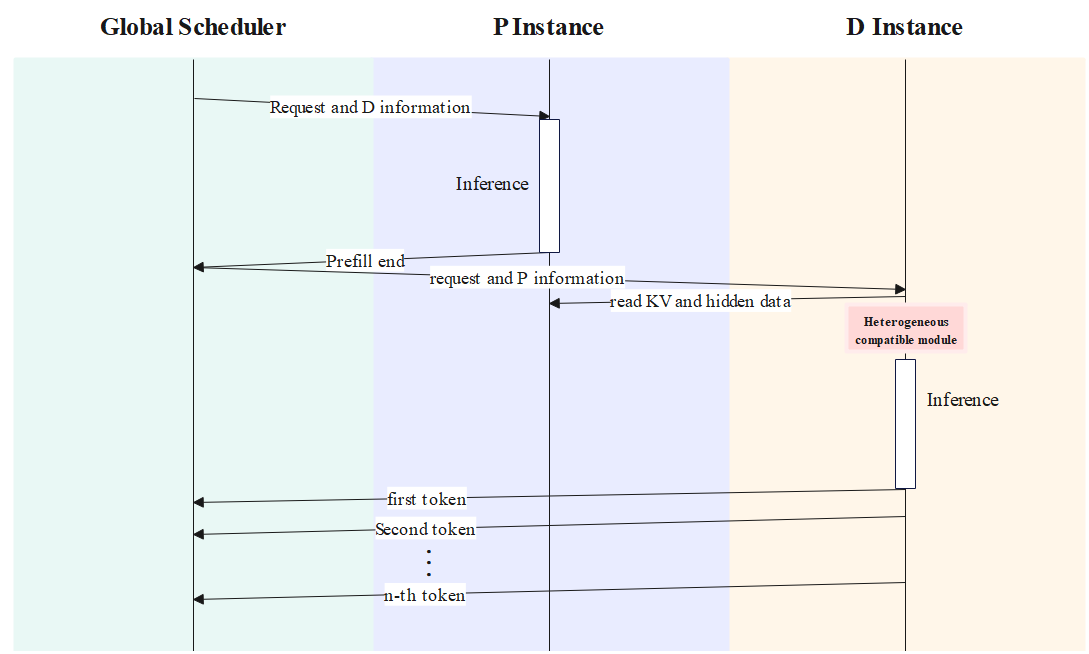}
	\caption{Workflow of P-D disaggregated Inference System}
	\label{Fig.2}
\end{figure}

The system workflow is shown in Fig.\ref{Fig.2}. First, the global scheduler receives the user request, and schedules the request to select the appropriate P instance and D instance according to the load of each instance. Second, the global scheduler sends the request that including the location information of D instance to the P instance. Third, P instance executes the request inference to generate the hidden state of the first token, and copies the kv data and the hidden data to the CPU buffer. Fourth, the global scheduler send the request to D instance. Fifth, D instance searches the kv data and the hidden data in the CPU buffer of P instance, and reads these data into the CPU buffer of D instance. Sixth, D instance loads these data into the GPU buffer to execute the inference of the remaining tokens. 

\subsection{Heterogeneous Compatible Transmission Module Design}
The heterogeneous compatible module is the core module of P-D separation system in heterogeneous GPU scenarios from different manufacturers. The module is used to mask the differences between different GPUs and ensure that GPU data can be shared between different GPUs. In order to achieve the objective, the most important problem to be solved is the alignment of the VRAM management of different GPUs. At the same time, in order to solve the heterogeneous parallel strategy problem caused by different GPU capabilities and P-D disaggregated characteristics, the parallel strategy alignment component is also embedded in the heterogeneous compatibility module. The detailed functional design of these three components is as follows.

\subsubsection{RDMA Transmission Module}

The kv data transmission between P instance and D instance is implemented through the read interface of the transfer engine. The input parameters of the transfer engine's read interface consist of three parts: the local buffer address, the remote buffer address and the remote location. These informations are obtained through control plane information interaction. The transfer engine component is self-developed, and the transfer engine of Mooncake open source project can be used. In order to reduce the occupancy of the HBM, the pinned memory of the CPU is allocated during the initialization process as a buffer to cache the kv data. The buffer is managed to reduce the overhead caused by temporary allocation, and is registered with RDMA to achieve zero-copy. 

\subsubsection{VRAM Management Alignment Component}

\begin{figure}
	\centering
	\includegraphics[width=8cm]{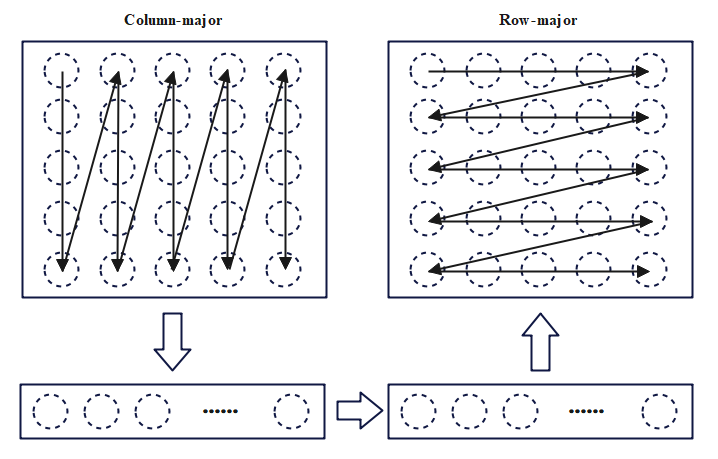}
	\caption{VRAM Management Alignment}
	\label{Fig.5}
\end{figure}

GPUs from different manufacturers may have different VRAM management methods, subject to the hardware architecture design, such as the block size of page attention and the layout of tensor data. Therefore, a VRAM management alignment component is designed to solve the above problems. The component can convert the layout of GPU data of P instance to execute the inference process of D instance according to the demand of D instance. A general method was designed, which converts a tensor into a one-dimensional tensor before transmission to eliminate the impact of layout, and then converts the one-dimensional tensor into the required memory layout after transmission. The layout conversion is shown in Fig.\ref{Fig.5}.

\subsubsection{Heterogeneous Parallel Strategy Alignment Component}

\begin{figure}
	\centering
	\includegraphics[width=8cm]{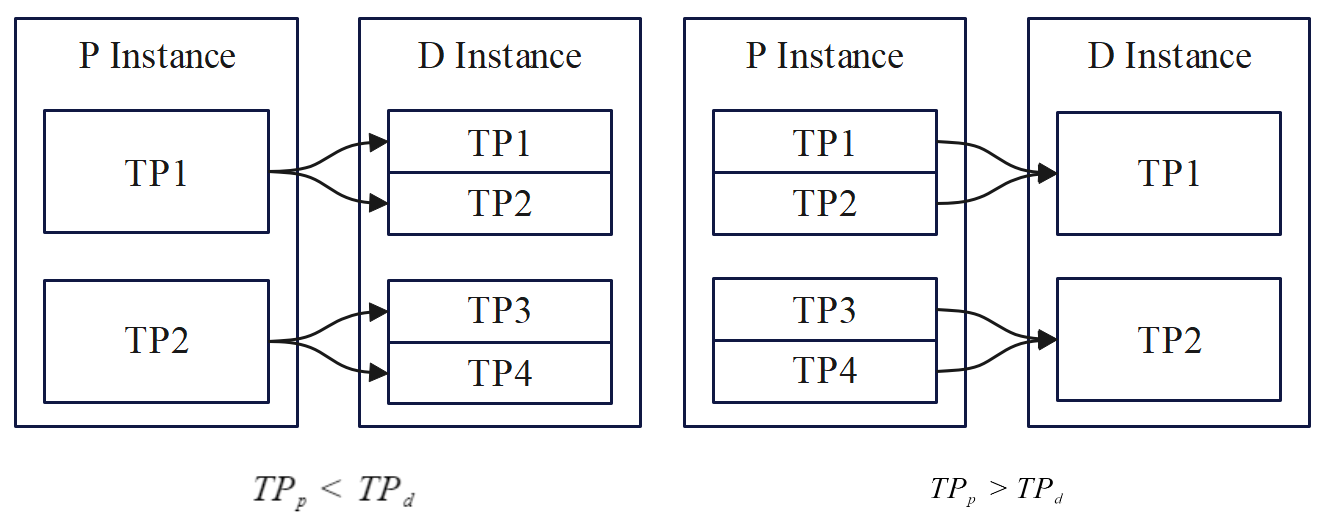}
	\caption{Heterogeneous Parallel Strategy Alignment}
	\label{Fig.4}
\end{figure}

In the P-D disaggregated inference system, P instances and D instances will adopt different parallel strategies in order to maximize the throughput of the inference system. Therefore, a heterogeneous parallel strategy alignment component is designed to combine or split GPU data for normal inference of D instances. The workflow is shown in Fig.\ref{Fig.2}. The request including the GPU ranks and parallel strategy of P instance is received by the D instance. Then, the D instance adopt the heterogeneous parallel compatible algorithm to reads the KV data from the P instance based on parallel strategies of P and D instances. For example, as shown in Fig.\ref{Fig.4}. If the TP=4 of P instance is greater than the TP=2 of D instance, then D instance needs to combine the TP1 and TP2 of P instance and transfer them to the TP1 of D instance, and combine the TP3 and TP4 of P instance to the TP2 of D instance. If the TP=2 of P instance is less than the TP=4 of D instance, the opposite splitting method is used.

\subsection{Joint Optimization of Parallel Strategy and P-D Ratio Algorithm}

After the P-D disaggregated inference system for heterogeneous GPUs from different manufacturers is established, a reasonable model deployment strategy is necessary based on customer scenario requirements. In response to QPS requirements of customer, the joint optimization problem of parallel strategy and P-D number ratio is proposed. Since the request first arrives the P instance and then arrives the D instance, the inference capability of P instance need to satisfy the QPS requirements, and the inference capability of D instance need to satisfy the inference capability of P instance. Therefore, the joint optimization problem is modeled as follows.

\subsubsection{P Instance Model}
The inference of P instance is the process that generates the first token. Let $T_p$ denotes the throughput of the P instance. $M_p$ denotes that the VRAM size of the GPU of P instance, and $m_p$ denotes the actual VRAM occupied size of P instance at a certain moment. $L_{ttft}$ denotes the threshold of the first generation time, and $l_p$ denotes the actual generation time of the first token. $dp_p$, $tp_p$, $pp_p$ and $ep_p$ denotes the data parallel, tensor parallel, pipeline parallel and expert parallel strategy of P instances, respectively.

Then, we provide a formulation of the problem of optimizing parallel strategies. The mathematical model with the objective of maximizing the throughput of each GPU, subject to the threshold of TTFT and VRAM size of GPU, is as follows:

\begin{equation}
	\label{P instance model}
	\begin{array}{l}
	\max \limits_{dp_p,tp_p,pp_p,ep_p} \frac{T^p}{dp_p*tp_p*pp_p} \\
	s.t.\\
	(c1)\;\; l_p(dp_p,tp_p,pp_p,ep_p) \leq L_{\emph{ttft}} \\
	(c2)\;\; m_p(dp_p,tp_p,pp_p,ep_p) \leq M_p \\
	\end{array}
\end{equation}

For the constraints, constraint (c1) indicates that the time for each P instance to generate the first token does not exceed the threshold $L_{\emph{ttft}}$. $tp_p$ denotes the parallel strategy of P instances. Constraint (c2) indicates that the VRAM occupancy of each GPU of each P instance does not exceed the VRAM of the GPU. The generation time of the first token $l_p$ mainly consists of two parts: computing time and communication time. In Eq.\eqref{P instance model}, $l_p$ can be calculated by 
\begin{equation}
	\label{l_p}
	l_p = \frac{c_{p}^{\emph{computing}}}{\lambda R_p} + \frac{e_p^{\emph{communication}}}{\beta B_p}
\end{equation}
where $c_p^{\emph{computing}}$, $e_p^{\emph{communication}}$ denote the computation and communication amount of P instance, respectively. $R_p$, $\lambda$ denote the computational capabilities and discount factor of P instance, respectively. $B_p$, $\beta$ denote the communication capability and discount factor of P instance, respectively. $m_p$ can be calculated by 

\begin{equation}
	m_p = m_p^{\emph{weight}}+m_p^{\emph{activation}}
\end{equation}
where $m_p^{\emph{weight}}$, $m_p^{\emph{activation}}$ denote the data size of weights and activations of P instance. 

\subsubsection{D Instance Model}
The inference of D instance is the process that generates remaining tokens. Let $T_y^d$ denotes the throughput of the $y$-th D instance. $Y$ denotes the number of D instances that need to be decided. $dp_d$, $tp_d$, $pp_d$ and $ep_d$ denotes the data parallel, tensor parallel, pipeline parallel and expert parallel strategy of Deep instances, respectively.

Then, we will provide the formulated problem of joint optimization of parallel strategy and instance number allocation. The mathematical model with the objective of maximizing the throughput of each D instance, subject to the threshold of TPOT and VRAM size of GPU, is as follows:
\begin{equation}
	\begin{array}{l}
		\max \limits_{dp_d,tp_d,pp_d,ep_d,Y} \frac{\sum\limits_y{T_y^d}}{Y} \\
		s.t. \\
		(c1)\;\; l_d(Y,dp_d,tp_d,pp_d,ep_d)\leq L_{\emph{tpot}} \\
		(c2)\;\; m_d(Y,dp_d,tp_d,pp_d,ep_d) \leq M_d
	\end{array}
\end{equation}

For the constraints, constraint (c1) indicates the time interval for each D instance to generate the remaining tokens does not exceed the threshold $L_{\emph{tpot}}$.

According to the characteristics of D instance, TPOT of D instance mainly consists of memory access time and communication time. The computation time of D instance is masked by the memory access time through operator design. Therefore, $l_d$ can be calculated by 

\begin{equation}
	l_d = \frac{e_d^{\emph{VRAM}}}{\alpha B_d^{\emph{VRAM}}} + \frac{e_d^{\emph{communication}}}{\beta B_d}
\end{equation}
where $e_d^{\emph{VRAM}}$, $e_d^{\emph{communication}}$ denote the data size of memory access and communication, respectively. $B_d^{\emph{VRAM}}$, $\alpha$ denote the memory access bandwidth and discount factor of D instance, respectively. $B_d$, $\beta$ denote the communication bandwidth and discount factor of D instance, respectively. The occupied VRAM of D instance mainly consists of three parts: weight, activation and KV. $m_d$ can be respresented by 

\begin{equation}
	m_d = m_d^{\emph{weight}} + m_d^{\emph{activation}} + m_d^{\emph{KV}}
\end{equation}

\subsection{Simulator Design}
\begin{figure}
	\centering
	\includegraphics[width=8cm]{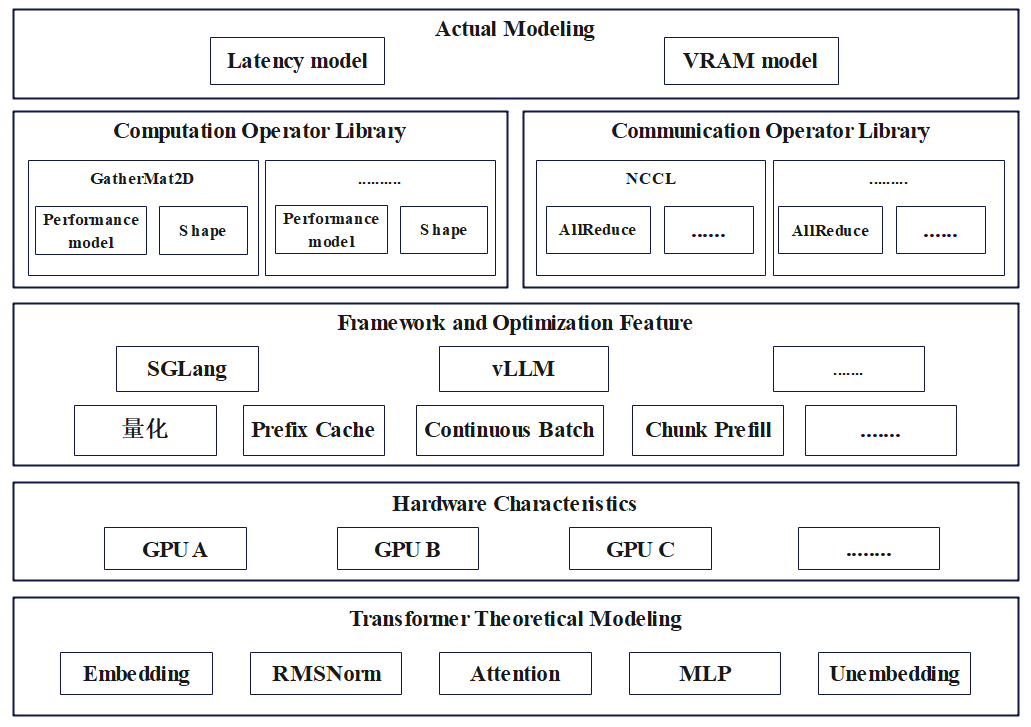}
	\caption{Simulator Model}
	\label{Fig.3}
\end{figure}

In order to solve the above joint optimization problem, we first need to obtain the values of each variable by modeling the inference process based on different models and hardwares. The overall modeling process is shown in Fig.\ref{Fig.3}, and the reasoning process is modeled in a hierarchical form. First, the bottom layer is the theoretical modeling of the transformer structure, without other constraints such as hardware and inference framework. The previous layer of the theoretical modeling layer is the hardware feature layer, which considers the relevant characteristics of the hardware on the basis of theoretical modeling, such as data alignment, memory management, etc. The previous layer of the hardware feature layer is the inference framework layer, which mainly needs to consider the optimization characteristics of each inference framework, such as prefix cache, quantization, etc. The previous layer of the framework layer is the operator library layer, which consists of two parts: computing operator library and communication operator library. The operator library is a collection of operators formed by the characteristics and constraints of hardware and framework. Based on the above work, the latency and VRAM usageof the inference system can be modeled to assist in optimizing the algorithm in the previous subsection. 

\section{Performance evaluation}
In the section, we evaluate the performance of inference system of Heterogeneous GPUs from different vendors. The experimental platform adopts GPUs from two vendors: GPU A (80G, 312TFLOPS) and GPU B (32G, 512TFLOPS). In the experimental, GPU B is used as P instance, and GPU A is used as D instance. Since the limited number of GPU, Llama2-7B is used as the experimental LLM. The specific experimental analysis is presented as follows. 

\subsection{Influence of Different Context Lengths}
\begin{figure*}
	\centering
	\subfloat[Latency(s)]{\includegraphics[width=6cm]{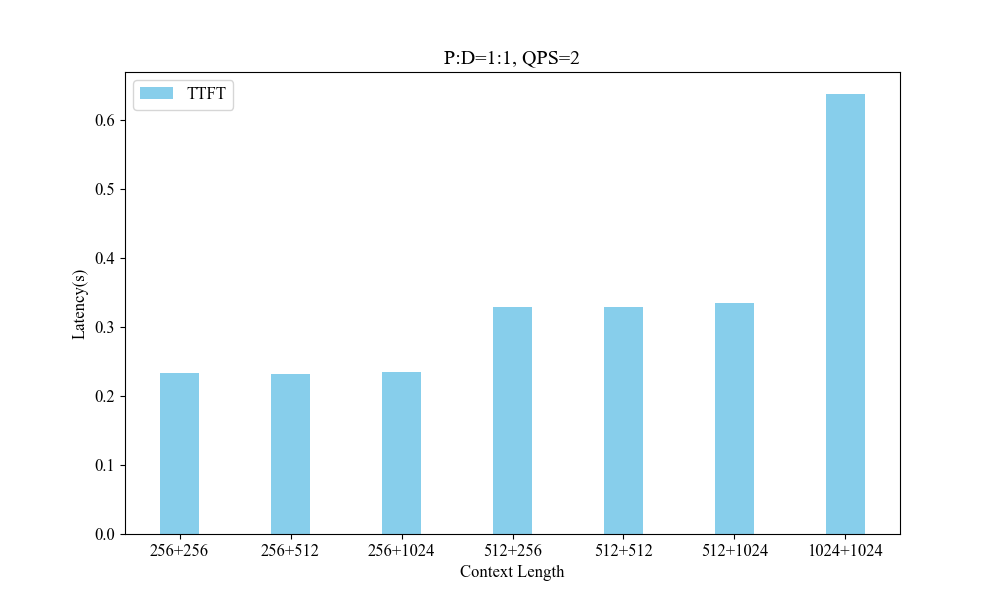}%
		\label{a}}
	\hfil
	\subfloat[Latency(s)]{\includegraphics[width=6cm]{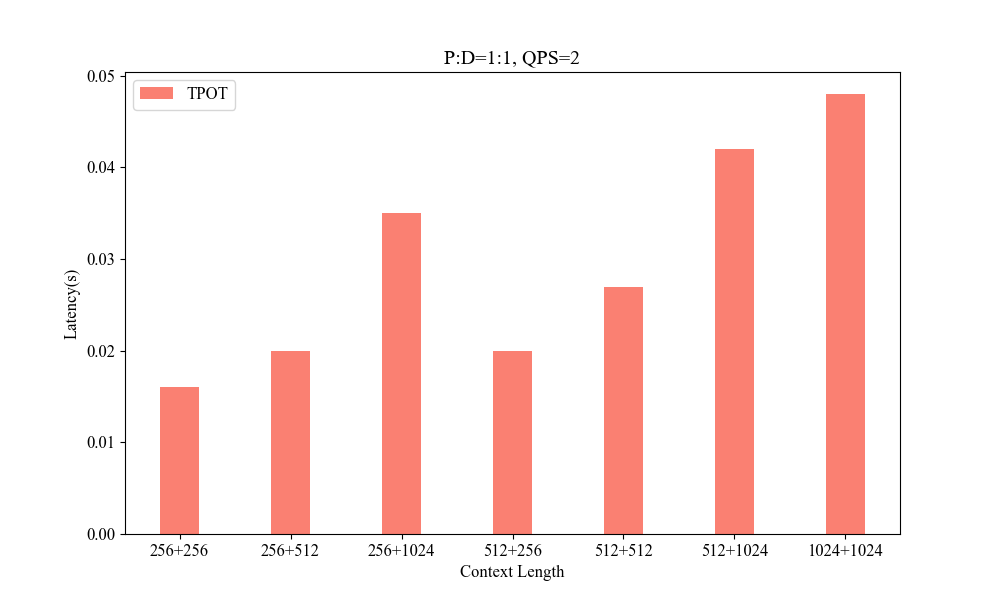}%
		\label{b}}
	\hfil
	\subfloat[Throughput(tokens/s)]{\includegraphics[width=6cm]{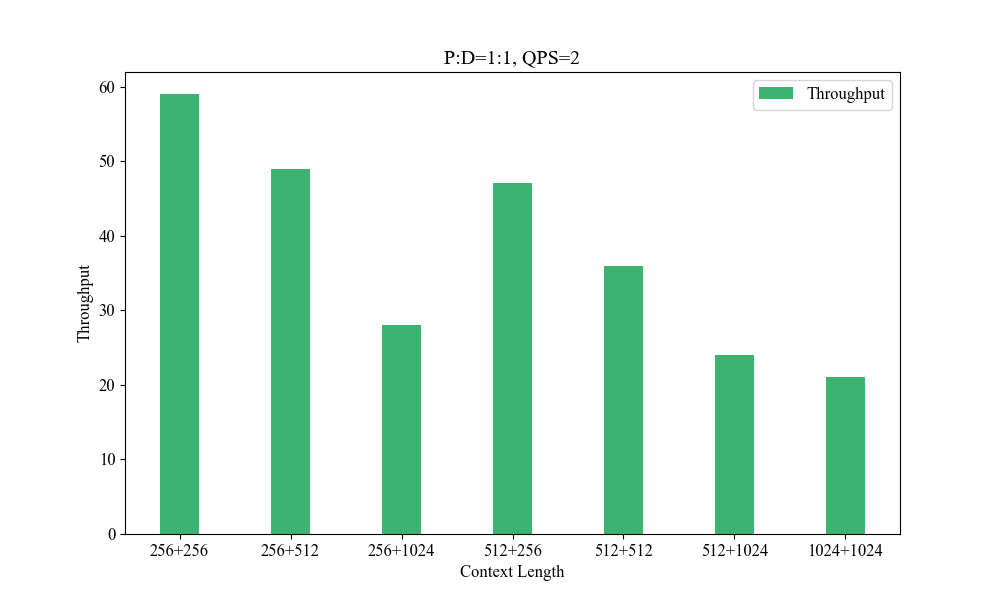}%
		\label{c}}
	\caption{Influence of Different Context Lengths}
	\label{Fig.6}
\end{figure*}

Figure \ref{Fig.6} shows the influence of different context lengths in the same configuration(P:D=1:1, QPS=2). From Figure \ref{Fig.6}(a), we observe that TTFT and TPOT will increase with the input and output lengths, and the TTFT time remains roughly the same in the same input length. But the TPOT time does not remains roughly the same in the same output length. Since the Decode stage is mainly memory-bound, it is related to the length of the input and output. From Figure \ref{Fig.6}(b), it is observed that as the length of input and output increases, the throughput decreases. 

\subsection{Influence of P-D Ratio}
\begin{figure*}
	\centering
	\subfloat[Latency(s)]{\includegraphics[width=6cm]{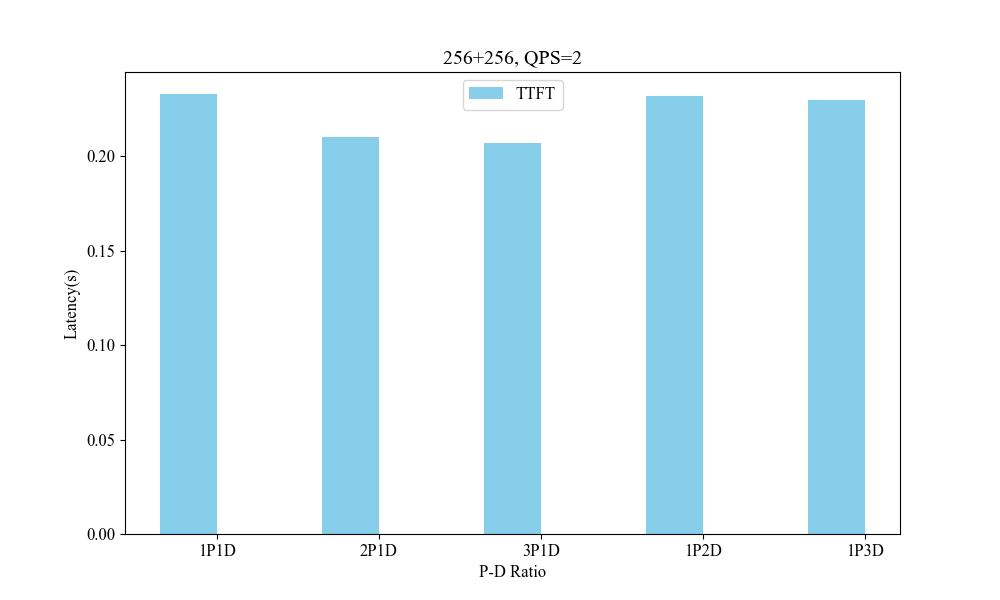}%
		\label{a}}
	\hfil
	\subfloat[Latency(s)]{\includegraphics[width=6cm]{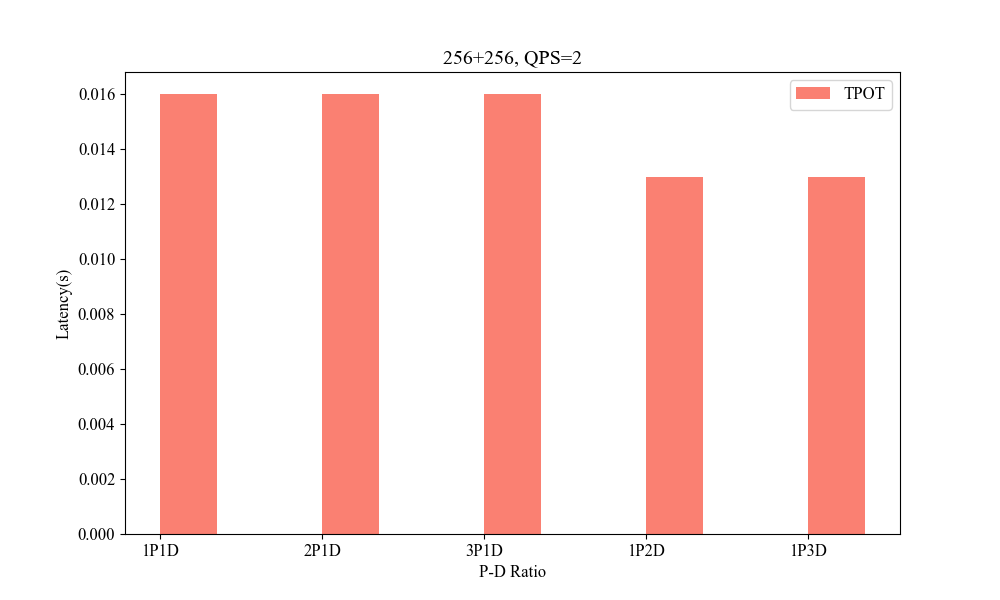}%
		\label{b}}
	\hfil
	\subfloat[Throughput(tokens/s)]{\includegraphics[width=6cm]{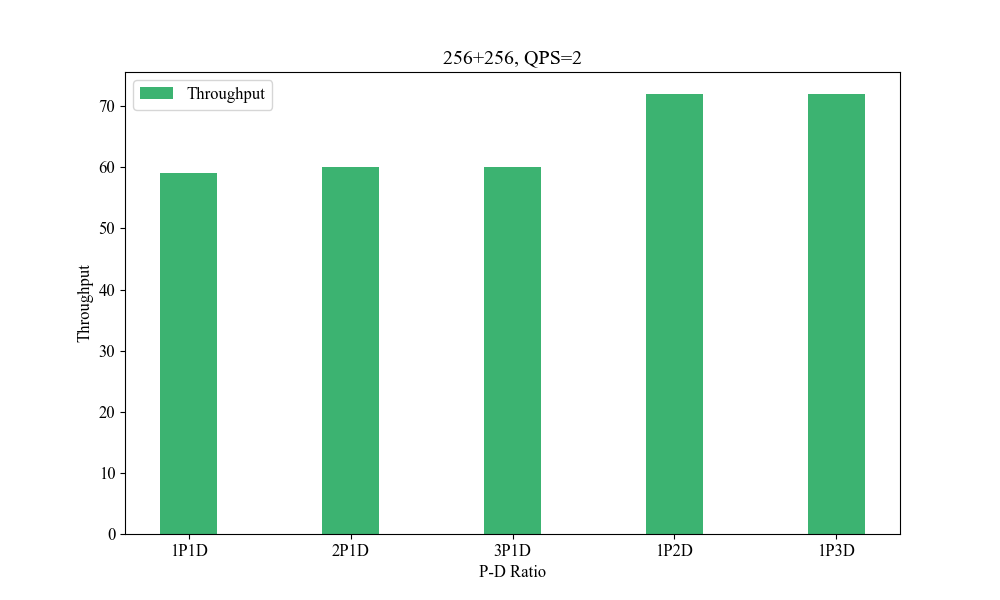}%
		\label{c}}
	\caption{Influence of P-D Ratio(256+256, QPS2)}
	\label{Fig.7}
\end{figure*}

\begin{figure*}
	\centering
	\subfloat[Latency(s)]{\includegraphics[width=6cm]{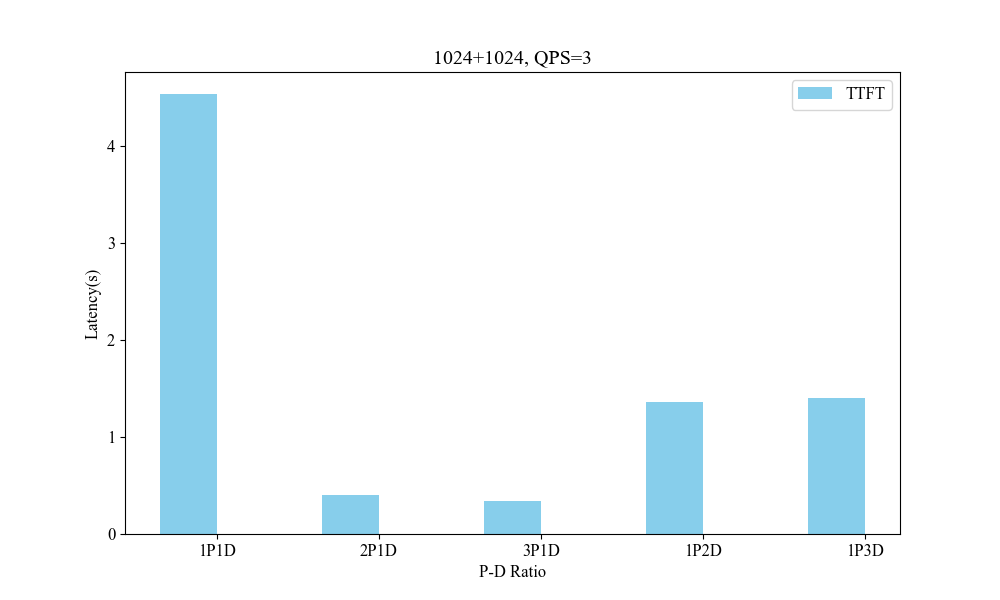}%
		\label{a}}
	\hfil
	\subfloat[Latency(s)]{\includegraphics[width=6cm]{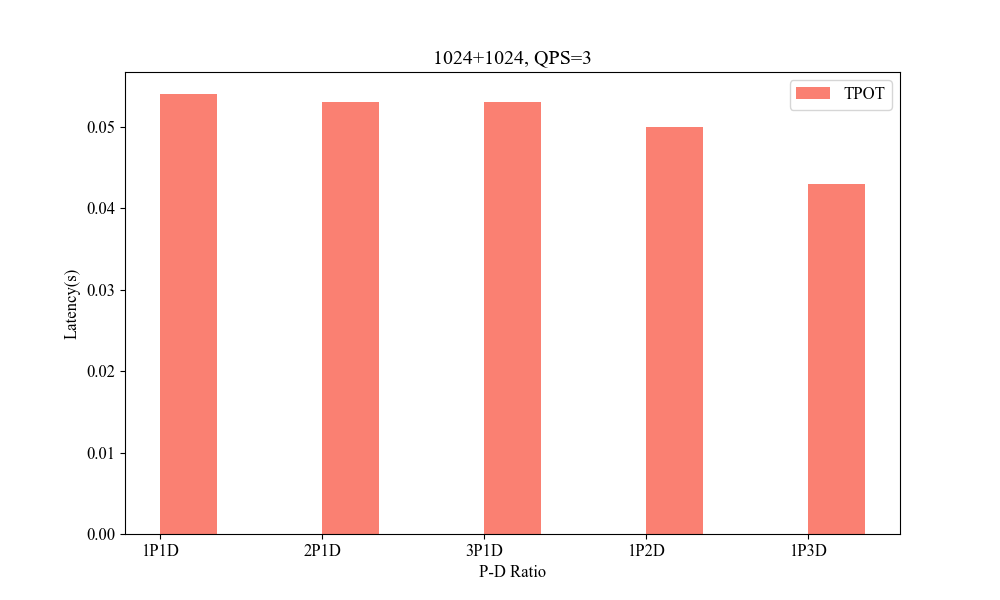}%
		\label{b}}
	\hfil
	\subfloat[Throughput(tokens/s)]{\includegraphics[width=6cm]{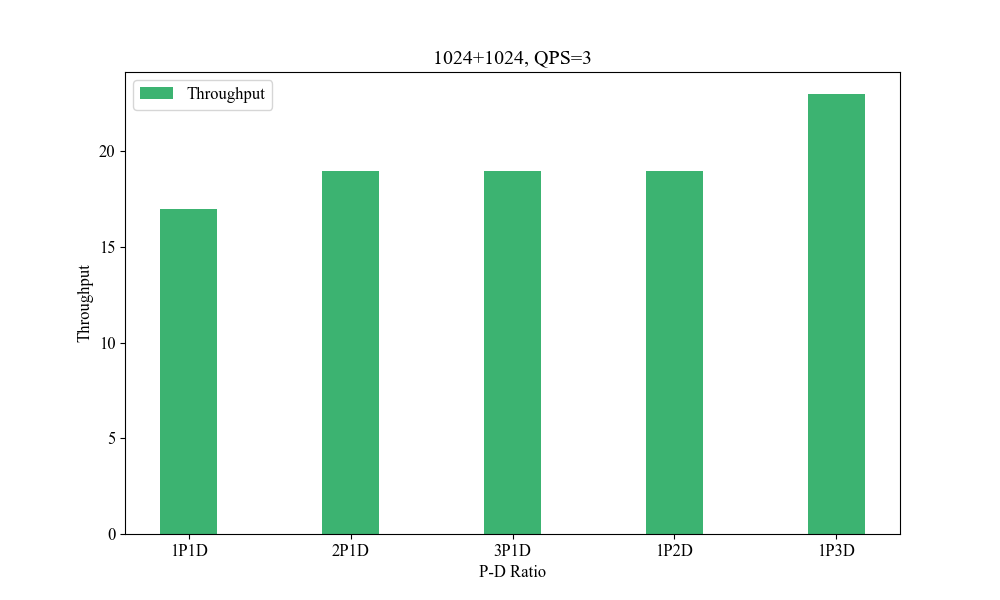}%
		\label{c}}
	\caption{Influence of P-D Ratio(1024+1024, QPS3)}
	\label{Fig.8}
\end{figure*}

Figure \ref{Fig.7} and \ref{Fig.8} show the influence of different P-D ratios(256+256, QPS2). From Figure \ref{Fig.7} (a), it can be observed that increasing the number of D instances can reduce the TPOT time, but it will not reduce exponentially (Since the decode computation utilization is low, increasing the D instance is not very beneficial in the case of short input and short output). At the same time, it can be observed that increasing the number of P instances can reduce the TTFT time, but it will not reduce exponentially (Since the input lengths and QPS are small, the P instance capabilities are not fully utilized, and increasing P instances does not bring much benefit). From Figure \ref{Fig.7} (b), it can be observed that 2P1D and 3P1D have a smaller throughput improvement than 1P1D, but the throughput of 2P1D and 3P1D is the same. Similarly, 1P2D and 1P3D have an improvement over 1P1D, but the throughput of 1P2D and 1P3D is the same. Therefore, Figure \ref{Fig.7} shows that the ratio between P and D is mutually constrained.

Figure \ref{Fig.8} shows the influence of different P-D ratios(1024+1024, QPS3). From Figure \ref{Fig.8} (a), it can be observed that increasing the number of D instances can reduce the TPOT time, but but it will not reduce exponentially(computation is not the performance bottleneck). At the same time, increasing the number of P instances can reduce the TTFT time, which will produce an exponential reduction phenomenon (when the input length, output length, and QPS are large, the P instances cannot handle the requests. Increasing the number of P instances has a greater benefit, which is suitable for the case where TTFT cannot meet the SLO).

\begin{figure*}
	\centering
	\subfloat[Latency(s)]{\includegraphics[width=6cm]{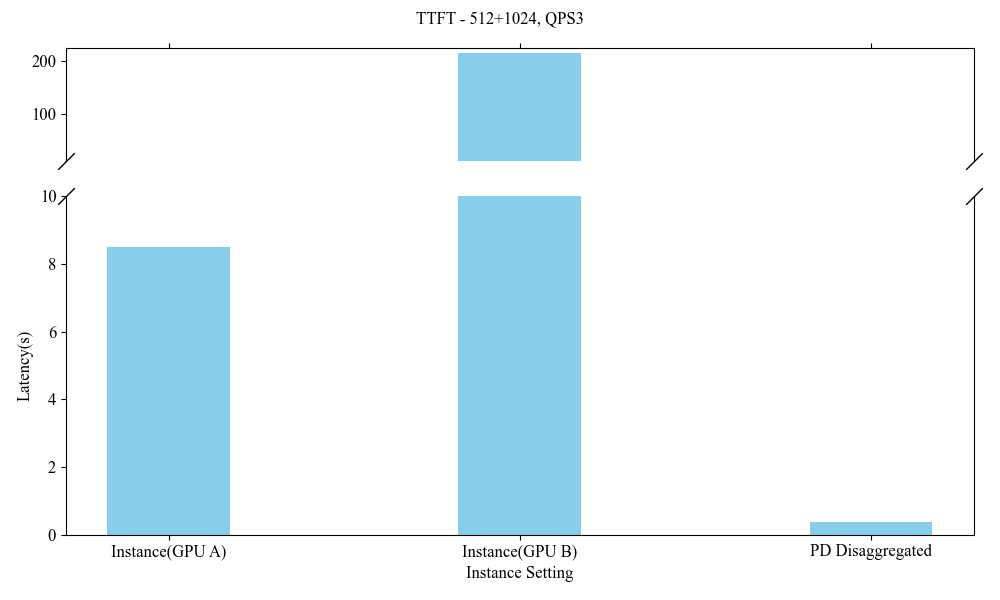}%
		\label{a}}
	\hfil
	\subfloat[Latency(s)]{\includegraphics[width=6cm]{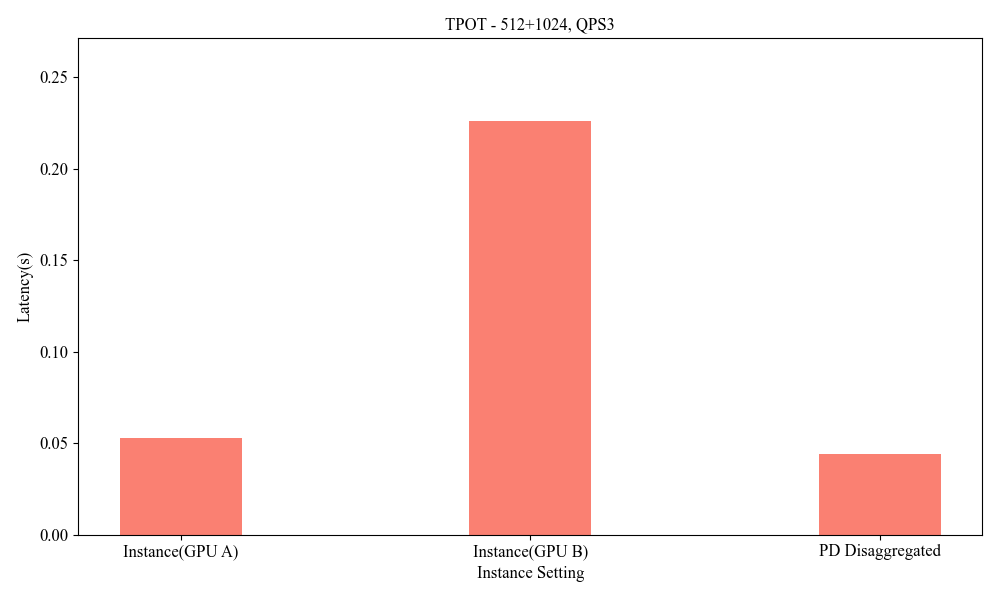}%
		\label{b}}
	\hfil
	\subfloat[Throughput(tokens/s)]{\includegraphics[width=6cm]{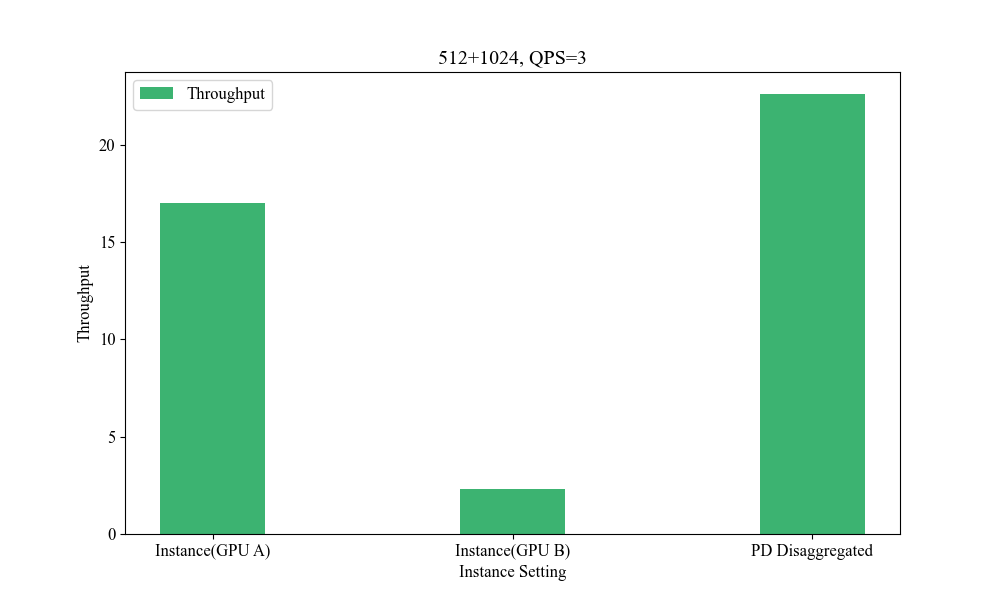}%
		\label{c}}
	\caption{Influence of Heterogeneous P-D(512+1024, QPS3)}
	\label{Fig.9}
\end{figure*}

\begin{figure*}
	\centering
	\subfloat[Latency(s)]{\includegraphics[width=6cm]{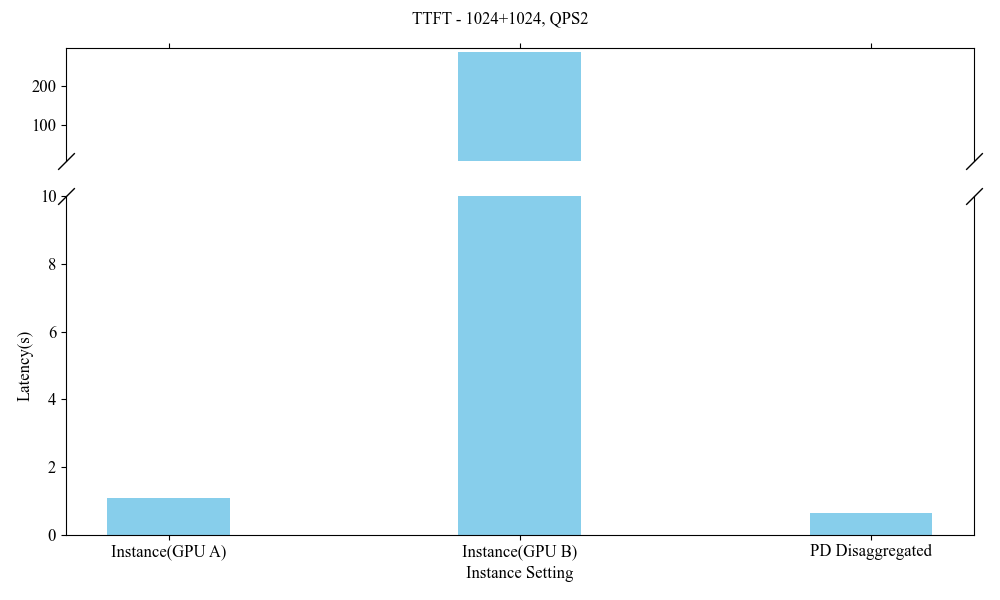}%
		\label{a}}
	\hfil
	\subfloat[Latency(s)]{\includegraphics[width=6cm]{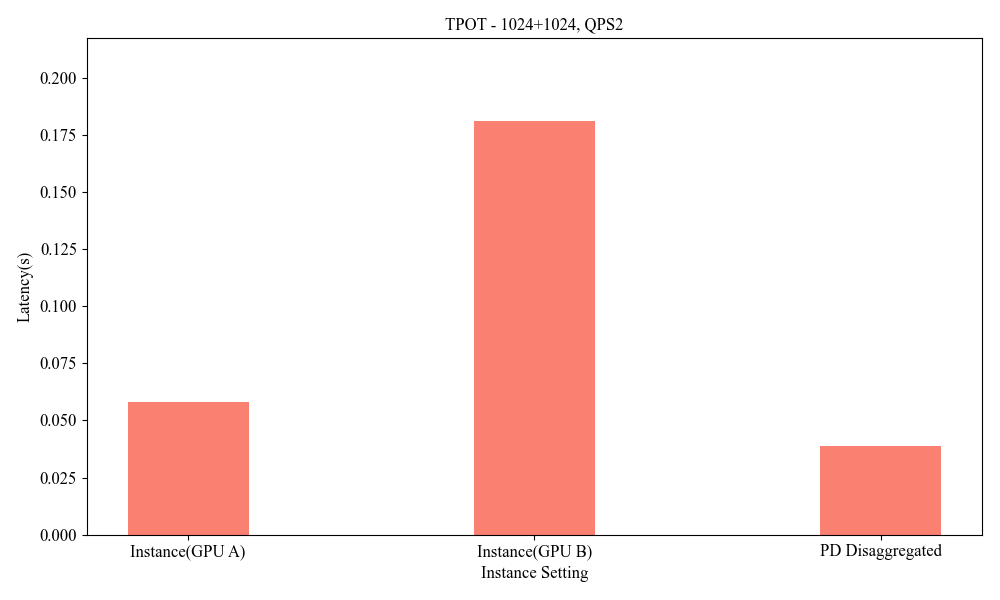}%
		\label{b}}
	\hfil
	\subfloat[Throughput(tokens/s)]{\includegraphics[width=6cm]{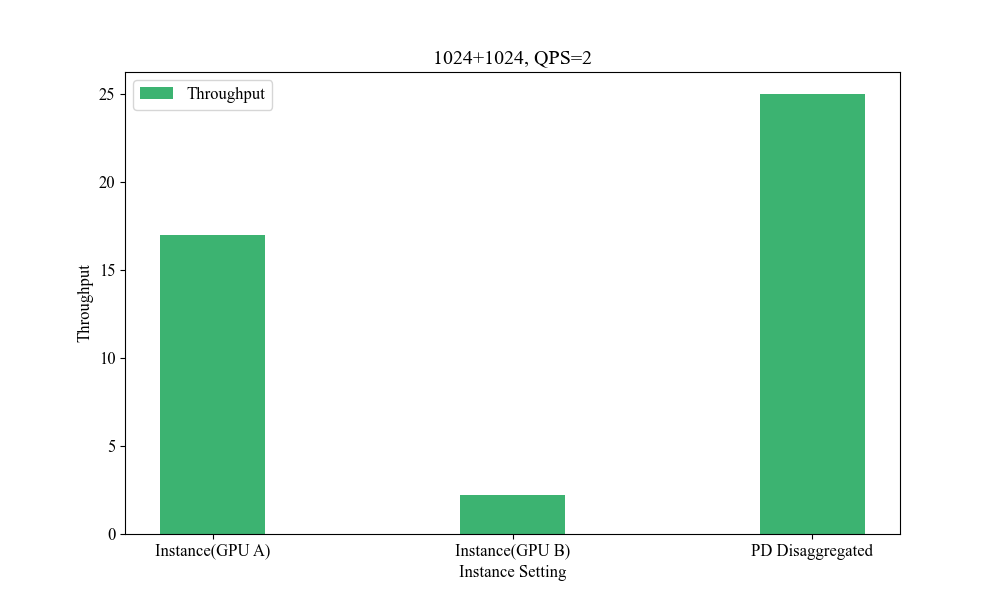}%
		\label{c}}
	\caption{Influence of Heterogeneous P-D(1024+1024, QPS2)}
	\label{Fig.10}
\end{figure*}
\subsection{Influence of Heterogeneous P-D}
Figure \ref{Fig.9} and \ref{Fig.10} show the influence of heterogeneous P-D in different context lengths and QPS. From Figure \ref{Fig.9} (a) and \ref{Fig.10} (a),  
it can be observed that the TTFT time does not meet the SLO constraint in the PD integration scenario (GPU A, 1024+1024 (QPS 2), 512+1024 (QPS 3)). However, the TTFT time does meet the SLO constraint in the PD disaggreagetd scenario (1024+1024 (QPS 2), 512+1024 (QPS 3)). Therefore, it can be concluded that P-D disaggreagetd can broaden the performance bottleneck of a single GPU. Meanwhile, as shown in Figure \ref{Fig.9} (b) and Figure \ref{Fig.10} (b), compared with PD integration scenarios, the throughput of PD disaggreagetd can be improved by 17\% (19.3-22.6) and 30\% (19.2-25) in the QPS3 512+1024 and QPS2 1024+1024, respectively. The above Figures show that in high QPS and long context scenarios, PD disaggreagetd can achieve better performance benefits than PD integration. Under the constraints of GPU capabilities, the higher the QPS and the longer the context, the higher the performance gain of PD disaggreagetd.

\subsection{Benefit Scenario of Heterogeneous P-D Disaggregated}
The benefit scenario of heterogeneous P-D disaggregated can be given based on the above analysis. First, if a GPU's processing power is weak and cannot meet the QPS and SLO requirements in the P-D integration scenario, it can be used in the P-D disaggreagetd scenario, and the system throughput can be improved by adjusting the P-D instance ratio. Second, P-D disaggreagetd can eliminate interference between prefill and decode, and improve resource utilization and reduce costs by selecting appropriate differentiated GPUs. Third, P-D disaggreagetd can maximize the processing capacity of the prefill and decode stages through a heterogeneous parallel strategy, allowing the prefill to handle long contexts in the case of TP and the decode to handle more requests in the case of DP, the subsection work is presented in the future.

\section{Conclusion}
In this paper, a P-D disaggreagetd heterogeneous inference framework is proposed to improve idle resource utilization and reduce costs. In this framework, a heterogeneous compatibility module is designed to solve the compatibility issues caused by different VRAM management and different parallel strategies of heterogeneous GPUs. Based on this, a joint optimization problem of parallel strategy and instance ratio was modeled and solved by building a detailed simulator. we provide numerical experimental results in terms of TTFT, TPOT and the throughput, which shows the heterogeneous P-D disaggreagetd has the better performance. Meanwhile, the benefit scenario of heterogeneous P-D disaggregated is given based on the experimental data. However, due to resource limitations, no experimental verification was performed on a larger model. In the future work, the Experiments will be continuously improved on larger models, and the kv transmission latency will be continuously optimized, such as the parallel transmission and calculation solution can be implemented through by-layer transmission.


%








\bibliographystyle{IEEEtran}
\bibliography{IEEEabrv,ref.bib}
\end{document}